\documentclass[pra,aps,twocolumn,10pt,showpacs,nofootinbib]{revtex4-1}
\usepackage[utf8]{inputenc}
\usepackage[T1]{fontenc}
\usepackage{amsmath}
\usepackage{amsfonts}
\usepackage{amssymb}
\usepackage{mathcmd}
\usepackage{graphicx} 
\usepackage{dsfont}
\usepackage{braket}
\usepackage{todonotes}
\usepackage{bm}
\usepackage{bbm}
\usepackage{mathrsfs}

\newcommand{\mi}{\mathrm{i}}
\newcommand{\me}{\mathrm{e}}

\begin{document}

\title{Highly-Degenerate Photonic Waveguide Structures for Holonomic Computation}
\author{Julien Pinske}
\author{Lucas Teuber}
\author{Stefan Scheel}
\email{stefan.scheel@uni-rostock.de}
\affiliation{Institut f\"ur Physik, Universit\"at Rostock, 
Albert-Einstein--Stra{\ss}e 23-24, D-18059 Rostock, Germany}

\date{\today}

\begin{abstract}
We investigate an all-out optical setup allowing for generation of non-Abelian 
geometric phases on its large degenerate eigenspaces. The proposal has the form 
of an $M$-pod system and can be implemented in terms of integrated photonic 
waveguide structures. We show that by injecting a larger number of photons into 
the optical setup, the degeneracy of eigenspaces scales rapidly. After studying 
the spectral properties of our system for the general case, we show how 
arbitrary $\mathrm{U}(4)$ transformations can be generated on the dark subspace 
of an optical tripod filled with two photons. 
Moreover, a degeneracy in the bright subspaces of the system, absent in 
any atomic analogue, allows for the generation of universal single-qubit 
manipulations. Finally, we address the complexity issue of holonomic 
computation. Particularly, we show how two-qubit and three-qubit states can be 
implemented on a photonic tripod, where a natural multi-partite structure is 
inherited from the spatial mode structure of the waveguides. 
\end{abstract}
\maketitle

\section{Introduction}
\label{sec:intro}

Quantum computation (QC) and quantum information processing are among the most 
promising developments in modern physics. Both subjects utilise the fact that 
the nonclassical nature of certain quantum systems allow for shortcuts in 
algorithmic evolutions, and in that way speeding up the computation, see
e.g.~\cite{Shor,Deutsch,Google}. Moreover, quantum information science proposes 
a number of results about the security of communication channels unmatched by 
any classical security protocol, e.g.~\cite{BB84,Gottesman}. However, the number 
of astonishing applications seems to be evenly matched by the number of 
technical challenges one encounters when faced with the task of building an 
actual quantum computer. Besides the patience and care an experimentalist can 
provide in preparing a stable and efficient experimental setup, there is an 
extensive literature on how to make QC robust and fault-tolerant against certain 
classes of errors \cite{Preskill,QEC}. 

An important subset of these techniques is referred to as topological quantum 
information \cite{Kitaev1,Kitaev2,TQC}. Roughly speaking, topological methods are 
based on the idea of error avoidance in contrast to error correction, i.e. 
protecting the quantum state from a decohering environment or parametric 
fluctuations, see e.g. Ref.~\cite{ZanarLloyd2}. In addition to this desirable 
symmetry-based protection of information, topological QC offers also a deeper 
insight into a number of geometric and topological notions at an experimentally 
feasible scale. These notions are not only central to much of modern 
mathematics, but are prominent features in theories of fundamental interactions 
as well. Therefore, there has been an increased interest in such topological 
systems with one focus being on the study and implementation of artificial gauge 
fields and symmetry groups \cite{Zoller1}. Results ranged from implementation 
of single artificial gauge fields in photonic \cite{Szameit2,Teuber} and atomic 
systems \cite{Spielman} to experimental simulation of lattice gauge 
theories \cite{Zoller2}.

In this article, we are mainly concerned with the paradigm of holonomic quantum 
computation (HQC) \cite{HQC,Pachos99}. HQC is a geometric approach to QC in 
which the manipulation of quantum information (qubits) is carried out by means 
of non-Abelian geometric phases following a closed parameter variation (quantum 
holonomies) \cite{Wilzeck}. It was shown in Ref.~\cite{HQC} that generically 
such a computation is universal. The advantage of constructing holonomic 
quantum gates lies in their parametric robustness so that, in principle, a 
family of (universal) fault-tolerant quantum gates can be designed in terms of 
holonomies only \cite{Oreshkov,Oreshkov2}. 

Besides its mathematical abstraction, the holonomic route to QC is associated 
with a series of technical challenges one has to overcome when implementing a 
(universal) holonomic quantum computer. In particular, HQC demands for the 
preparation of large degenerate eigenspaces which act as a quantum 
code~\cite{Laflamme}. To be precise, for a code $\mathcal{C}$ consisting of 
$k$-qubit code words we need an eigenspace of dimension of at least $2^{k}$.  
Usually, the degeneracy is ensured by some form of symmetry, i.e. code words in 
certain subspaces of $\mathcal{C}$ cannot be distinguished energetically.
The preparation of such highly symmetric quantum codes becomes a demanding 
experimental challenge as $k$ increases.

A typical implementation of geometric phases utilises $M$-pod systems 
\cite{Recati} that have their origin in atomic physics \cite{Bergmann}.
There, a collection of $M$ ground states is individually coupled to one excited 
state. Adiabatic parameter variations of the couplings that return to the 
initial configuration can then implement a (non-Abelian) geometric phase on the 
$(M-1)$-fold degenerate dark subspace. For instance, a realisation with trapped 
ions for the case of a tripod ($M=3$) was suggested in Ref.~\cite{Duan}, and 
another implementation utilised semiconductor macroatoms \cite{Rossi}.
As an extension, there also exist schemes with nonadiabatic parameter 
variations \cite{Sloeqvist,Abdumalikov,Kosaka,Peng}. The drawbacks of these 
systems are that an increase in degeneracy in associated with an increase in the 
number of ground states, whose implementation can become a challenging 
task. Furthermore, the listed proposals do not give rise to a proper 
multi-partite structure by themselves, and many-body interactions have to be 
considered to construct more intriguing gates.

Here, we present a linear optical implementation of the $M$-pod in the 
$N$-photon Fock layer giving rise to arbitrarily large degenerate eigenspaces on 
which HQC could be based. Our proposal can be realised solely in terms of 
integrated photonic structures such as laser-written silica-based 
waveguides \cite{Szameit}. The latter has been proven to be a versatile tool 
box that combines the proven capabilities of modern quantum optics,
like quantum communication \cite{Gisin}, implementing quantum 
devices \cite{Weinfurter} and gates \cite{Mataloni,Benson},
or initialising nonclassical states of light \cite{Moss,Tang}, with a high 
degree of interferometric stability. Hence, combining the coherence preserving 
properties of such structures with the intrinsic robustness of topological QC is 
a desirable aim.

In fact, there already exist a number of sophisticated works on optical 
holonomic quantum computation. A first theoretical proposal for an optical 
holonomic quantum computer goes back to Ref.~\cite{PachosOpt}, where a quantum 
holonomy generated from a nonlinear Kerr Hamiltonian was designed by driving 
squeezing and displacement in a suitable manner such that the desired gate can 
be obtained. In comparison to this early idea, our proposal is solely based on 
linear evanescent coupling of the waveguides and thus, nonlinearities have to 
be added to extend optical HQC to a universal scheme 
\cite{Shen}, e.g. measurement-based~\cite{KLM,Scheel}.
In a more recent work on optical HQC, spin-orbit coupling of polarised light in 
asymmetric microcavities is utilised to generate a geometric phase 
\cite{Schmidt}, whereas the emergence of a non-Abelian Berry phase was observed 
when injecting coherent states of light into topologically guided modes 
\cite{Chamon}. In Ref.~\cite{Teuber}, an artificial non-Abelian gauge potential 
was designed by driving an adiabatic path in the dark subspace of an optical 
tripod. However, the current implementations are all limited to only doubly
degenerate subspaces and thus only enable the study of $\mathrm{U}(2)$ 
holonomies.

In our present work, we overcome the limitations of current HQC schemes by 
increasing the number of photons involved in the dynamical evolution. As a 
result, the degeneracy of the system is easily increased. We further propose 
that, from the large number of degenerate eigenspaces, one will eventually 
find a subspace on which the spatial mode structure of the waveguide modes can 
be used to label logical qubits, even though the entire eigenspace might not 
possess a proper multi-partite structure. Therefore, our proposal overcomes 
another problem frequently occurring in HQC since its inception \cite{Pachos99}.

We finally note that, as a linear optical scheme, our setup is closely related 
to standard approaches of linear optical quantum computation based on large 
networks of beam splitters \cite{Reck-Zeil,O'Brien,Laing}. However, in 
constrast to such schemes, where each beam splitter (and phase shifter) has to 
be adjusted individually with potential fluctuations, holonomic approaches 
generate the desired evolution in one collective fault-tolerant dynamic which is 
especially true if larger degeneracies are achieved.

The structure of the article is as follows. In Sec.~\ref{sec:degeneracy}, we 
present the quantum optical $M$-pod which allows for the implementation of 
arbitrarily large degenerate subspaces and study its spectral properties.
Section~\ref{sec:HQC} is dedicated to a review of the basic theory of HQC, to 
the extent as it is relevant to our study of the photonic setup.
In Sec.~\ref{sec:example}, a detailed discussion of an optical tripod,
represented in the two-photon Fock layer, illustrates how arbitrary
holonomic $\mathrm{U}(4)$ transformations can be realised within the degenerate 
eigenspaces of the waveguide arrangement. 
The complexity question regarding the construction of logical quantum 
information (code words) is addressed in Sec.~\ref{sec:complexity}, where we 
show how to define the code words on the total Hilbert space of the system 
and act with holonomies from different eigenspaces onto a code $\mathcal{C}$. 
More precisely, we show how one can use the optical tripod to prepare 
two-qubit and three-qubit states. Finally, Sec.~\ref{sec:conclusions} contains 
a summary of the article as well as some concluding remarks.
In Appendix~\ref{app:DS&BS} we give an explicit parameter-dependent representation of the dark and bright states in the two-photon Fock layer. 
App.~\ref{app:subSpaceDim} contains details of a diagonalisation of the 
$M$-pod system in terms of bosonic field operators, showing how the eigenstates 
distribute over the different eigenspaces. In App.~\ref{app:loops} we design 
simple parameter variations that induce a number of useful quantum 
gates via a non-Abelian geometric phase. 

\section{Degeneracy in photonic waveguides}
\label{sec:degeneracy}

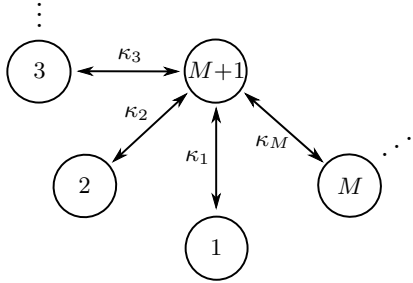
\begin{figure}[h]
\begin{tikzpicture}
\node at (0,0) {\includegraphics[width=5cm]{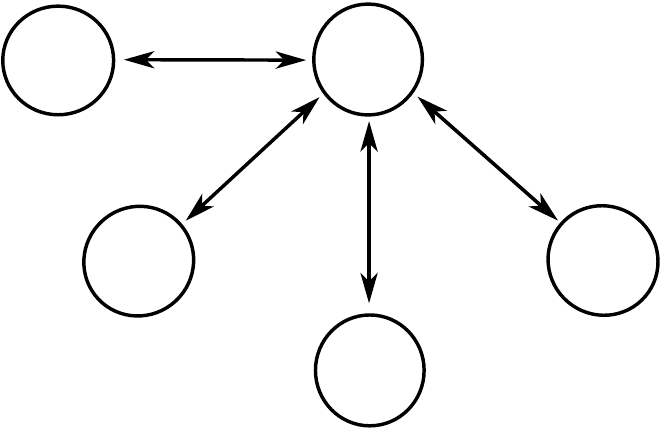}}; \node at (-2.05,2.05) 
{\large{$\vdots$}}; \node at (2.7,0.3) {\reflectbox{\large{$\ddots$}}};
\node at (0.3,1.2) {$M\hspace{-0.07cm}+\hspace{-0.07cm}1$}; \node at 
(-2.05,1.2) {$3$}; \node at (-1.45,-0.33) {$2$}; \node at (0.3,-1.15) {$1$}; 
\node at (2.08,-0.33) {$M$};
\node at (-0.85,1.4) {$\kappa_{3}$}; \node at (-0.76,0.63) {$\kappa_{2}$}; 
\node at (0.05,0.05) {$\kappa_{1}$}; \node at (1.05,0.25) {$\kappa_{M}$};
\node at (0,2.6) {\textcolor{white}{Hello world!}};
\end{tikzpicture}
\caption{\label{fig:M-pod} Schematic front view of an $M$-pod waveguide 
arrangement subject to the Hamiltonian in Eq.~(\ref{eq:M-pod}). 
The $i$-th outer waveguide ($i=1,\dots,M$) interacts solely with the central 
one via the coupling $\kappa_{i}$.}
\end{figure}

Let us consider the optical setup depicted in Fig.~\ref{fig:M-pod}, in which 
$M+1$ waveguides are arranged in the form of an $M$-pod. The Hamiltonian of the 
system reads 
\begin{equation}
\label{eq:M-pod}
H=\sum_{i=1}^{M}
\left(\kappa_{i}a_{i}a^{\dagger}_{M+1}+\kappa_{i}^{*}a_{i}^{\dagger}a_{M+1}
\right)
\end{equation}
(we set $\hbar=1$ throughout this work), where $a_{j}^\dagger$ ($a_{j}$) is the 
bosonic creation (annihilation) operator for the $j$-th waveguide mode 
($j=1,\dots,M+1$) and
$\kappa_{i}$ is the coupling strength between the $i$-th outer waveguide with 
the central waveguide. 

Next, we restrict the Hamiltonian~(\ref{eq:M-pod}) to act solely on the 
$N$-photon Fock layer  
\begin{equation}
 \mathcal{F}_{N}=\biggr\{\ket{n_{1},\dots 
n_{M+1}}\,\biggr\vert\,\sum_{j=1}^{M+1}n_{j}=N\biggr\}.
\end{equation}
Represented in the basis $\mathcal{F}_{N}$, the Hamiltonian $H$ from 
Eq.~(\ref{eq:M-pod}) defines an operator on the reduced Hilbert space
$\mathcal{H}=\mathrm{Span}(\mathcal{F}_{N})$ having dimension 
$p=(N+M)!/(N!M!)$, which is the number of possibilities to distribute
$N$ identical photons on $(M+1)$ labelled waveguides.

Choosing an index system for the $p$ Fock states, one can calculate a matrix 
representation $\bm{H}=(H_{ij})_{i,j=1}^{p}$ in $\mathcal{F}_{N}$. By 
diagonalisation of this matrix one finds a decomposition of the Hilbert space 
$\mathcal{H}$ into orthogonal eigenspaces, viz.
\begin{equation}
\label{eq:decomp}
\mathcal{H}=\mathcal{H}_{\mathcal{D}}\oplus\mathcal{H}_{\mathcal{B}_{+}}
\oplus\mathcal{H}_{\mathcal{B}_{-}}\oplus\dots\oplus
\mathcal{H}_{\mathcal{B}_{+}}^{(N)}\oplus\mathcal{H}_{\mathcal{B}_{-}}^{(N)},
\end{equation}
where $\mathcal{H}_{\mathcal{D}}$ is its dark subspace (eigenspace with
eigenvalue zero), and $\mathcal{H}_{\mathcal{B}_{\pm}}^{(n)}$ is 
the eigenspace corresponding to the energy $\pm n\varepsilon=\pm 
n\sqrt{\vert\kappa_{1}\vert^2+\dots+\vert\kappa_{M}\vert^2}$ ($n=1,\dots,N$).

\begin{figure}[h]
\centering
\includegraphics[width=0.49\textwidth]{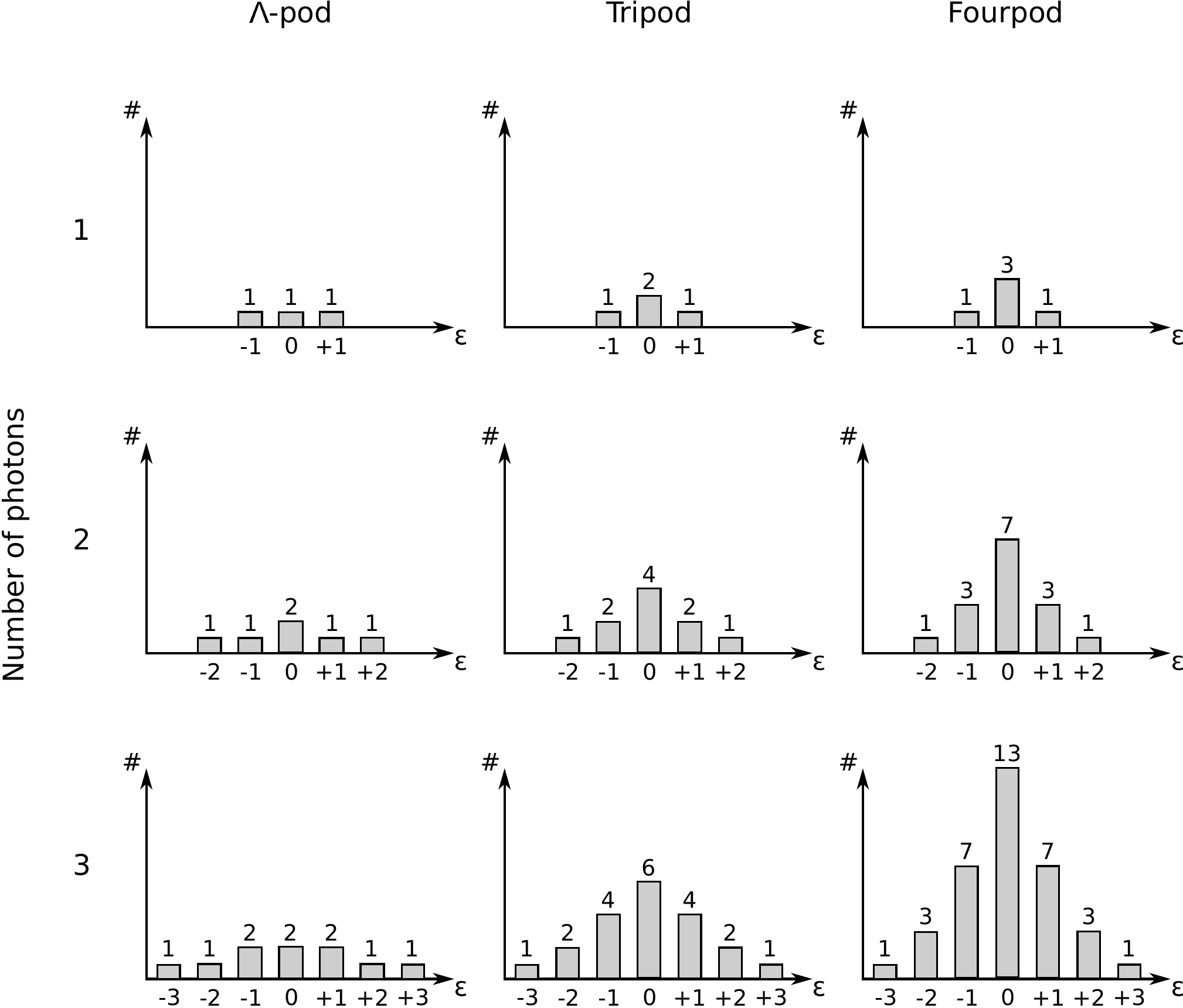}
\caption{\label{fig:spectral} Spectrum of an $M$-pod system filled with $N$ photons. 
The number of eigenstates ($\#$) is depicted over the energy (in multiples of $\varepsilon$) for $M,N\in\{1,2,3\}$.  
Additional waveguides lead to an increase in the degree of degeneracy of the eigenenergies. 
The degeneracy of the lowest levels (magnitudewise) increases the most, while the two highest (magnitudewise) energy levels are nondegenerate. 
In comparison, increasing the number of photons in the system, not only increases degeneracy but gives rise to two additional nondegenerate eigenenergies.
The profile of the bar diagram becomes steeper as $M$ increases.}
\end{figure}

The degeneracy of these subspaces depends on the number of waveguides and 
photons and can explicitly be calculated, see App.~\ref{app:subSpaceDim}.
For the dimension $d(N,M)$ of $\mathcal{H}_{\mathcal{D}}$ we find
\begin{align}
 d(N,M) = \begin{cases}
\sum_{n=1}^{N/2} { {2 n + M - 2} \choose {2 n}} , & \text{if $N$ even,} \\
\sum_{n=0}^{(N-1)/2} { {2 n + 1 + M - 2} \choose {2 n + 1}}, & \text{if $N$ 
odd,}
\end{cases}
\end{align}
and for all subsequent subspaces $\mathcal{H}_{\mathcal{B}_{\pm}}, 
\mathcal{H}_{\mathcal{B}_{\pm}}^{(2)}, \dots$ one finds $d(N-1,M), d(N-2,M), 
\dots$, respectively.
In Fig.~\ref{fig:spectral}, the resulting spectral structure is schematically 
shown for selected values of $N$ and $M$. Clearly, the addition of more photons 
drastically increases the available dimensions of subspaces on which one can 
perform HQC protocols.

In comparison, adding another state to an atomic $M$-pod system (which might be 
a challenging task) yields only one additional dark state. By utilising the 
tools of waveguide quantum optics one is thus able to increase the degeneracy 
in two ways. First, one can engineer an additional waveguide with coupling 
solely to the central waveguide (see Fig.~\ref{fig:M-pod}). This will increase 
the dimension of each eigenspace depending on the number of photons 
participating in the optical experiment. From Fig.~\ref{fig:M-pod}, one observes
that increasing the number of waveguide arms in the $M$-pod will fail when $M$ 
becomes large, because placing too many waveguides around the central one will 
ultimately result in a coupling of the outer waveguides with each other, thus 
breaking the structure of the $M$-pod. This problem can be avoided by following 
the alternative route by sending a higher number of photons into the $M$-pod. 

In the following, after reviewing some basic theory, we will illustrate an 
interesting application, in which the prepared degenerate eigenspaces are 
utilised to allow for the manipulation of (geometric) quantum information.
It turns out that our photonic setup allows, in principle, to implement 
rotations between the degenerate eigenstates such that the whole unitary group 
$\mathrm{U}(m)$ for arbitrary $m\in\mathbb{N}$ can be spanned in a purely 
geometric way.

\section{Computation with holonomies}
\label{sec:HQC}

For a working HQC procedure one seeks a computational scheme in which a 
geometric property, the holonomy, plays the role of the unitary gate. 
For the following investigation we make the usual assumption that the 
Hamiltonian $H$ of a quantum system can be expressed in terms of control fields 
$\bm{\lambda}=\{\lambda_{\mu}\}_{\mu=1}^{D}$ (couplings) 
which serve as local coordinates on a $D$-dimensional parameter space 
$\mathcal{M}$ (control manifold). 
If one is able to drive the control field configuration through a (piecewise) 
smooth path $\gamma:[0,T]\rightarrow \mathcal{M}$, we have 
$H[\bm{\lambda}(t)]=H_{\gamma(t)}$ and the quantum system evolves according to 
$U(T)=\boldsymbol{\hat{\mathrm{T}}}\me^{-\mi\int_{0}^{T}H_{\gamma(t)}\mathrm{d}t
}$ ($\boldsymbol{\hat{\mathrm{T}}}$ denotes time ordering). In this context, 
HQC is based on the idea that generating a sufficient, finite set of paths 
$\{\gamma_{i}\}$ induces a sequence of corresponding gates $\{U_{i}\}$ 
implementing the whole quantum information network \cite{HQC}.

Here, we are not interested in arbitrary paths but in those that represent 
loops in $\mathcal{M}$, that is, $\gamma(T)=\gamma(0)=\bm{\lambda}_{0}$. 
Let us further suppose that the Hamiltonian defines an iso-degenerate (no 
level-crossing) family of Hermitian operators with $R$ different eigenvalues. 
Then one has the $\bm{\lambda}$-dependent spectral 
decomposition $H(\bm{\lambda})=\sum_{j=1}^{R}\varepsilon_{j}(\bm{\lambda})\Pi_{j}(\bm{\lambda})$.
Here $\Pi_{j}(\bm{\lambda})$ is the projector onto the $m_{j}$-fold degenerate 
eigenspace $\mathcal{H}_{j}(\bm{\lambda})=
\mathrm{Span}(\{\ket{\psi_{j,a}(\bm{\lambda})}\}^ {m_{j}}_{a=1})$,
corresponding to the energy $\varepsilon_{j}(\bm{\lambda})$.

We restrict ourselves to adiabatic loops, i.e. the change of the control fields 
happens slow enough such that transitions to states of different eigenenergies 
are prohibited \cite{Fock}. 
From the adiabatic assumption it follows that any initial preparation 
$\ket{\psi(0)}=\ket{\psi_{\mathrm{in}}}\in\mathcal{H}_{j}$ is mapped,
after a time period $T$, onto a final state $U(T)\ket{\psi_{\mathrm{in}}}$ 
lying also in $\mathcal{H}_{j}$. Hence, the time evolution consists of a sum of 
unitary evolutions within each degenerate subspace $\mathcal{H}_{j}$. 
Explicitly, we have \cite{HQC}
\begin{eqnarray}
U(T) & = & \oplus_{j=1}^{R}\me^{\mi\omega_{j}(T)}W^{(j)}(\gamma), 
\label{eq:full-holo}\\
W^{(j)}(\gamma) & = & \boldsymbol{\hat{\mathrm{P}}}
\mathrm{exp}\oint_{\gamma}A_{j}\in\mathrm{U}(m_{j}),\label{eq:holo}
\end{eqnarray}
where the first exponential, with 
$\omega_{j}(T)=\int_{0}^{T}\varepsilon_{j}(\bm{\lambda}(t))\mathrm{d}t$, 
is the dynamical phase while the second term $W^{(j)}(\gamma)$ is a quantum 
holonomy determined by the path-ordered exponentiation of the matrix-valued 
phase factor $\oint A_{j}$. Here, 
$A_{j}=\Pi_{j}(\bm{\lambda})\mathrm{d}\Pi_{j}(\bm{\lambda})$ is a non-Abelian 
gauge potential often referred to as the adiabatic connection (local connection 
one-form). Its (anti-Hermitian) components read 
\begin{equation}
\label{eq:gauge}
 \left(A_{j,\mu}\right)_{ab}=
 \bra {\psi_{j,a}(\bm{\lambda})}\partial_{\mu}\ket{\psi_{j,b}(\bm{\lambda})},
\end{equation}
such that $A_{j}=\sum_{\mu=1}^{D}A_{j,\mu}\mathrm{d}\lambda_{\mu}$ 
($\partial_{\mu}=\partial/\partial\lambda_{\mu}$). 

The set of transformations $W^{(j)}(\{\gamma_{i}\})$ generated from a set of 
loops $\{\gamma_{i}\}$ at an initial point $\bm{\lambda}_{0}$ forms a subgroup 
of the unitary group $\mathrm{U}(m_{j})$ that is known as the holonomy group 
$\mathrm{Hol}(A_{j})$. A lower bound for the dimension of $\mathrm{Hol}(A_{j})$ 
is given by the number of linear independent components of the local curvature 
two-form 
$F_{j}=\sum_{\mu<\nu}^{D}F_{j,\mu\nu}\mathrm{d}\lambda_{\mu}\wedge\mathrm{d}
\lambda_{\nu}$ ($\wedge$ denotes the antisymmetrised tensor product) 
\cite{Recati,Nakahara}. Its antisymmetric components can be computed from 
\begin{equation}
 \label{eq:curvature}
 F_{j,\mu\nu}=\partial_{\mu}A_{j,\nu}-\partial_{\nu}A_{j,\mu}+[A_{j,\mu},A_{j,\nu}],
\end{equation}
where $[A_{j,\mu},A_{j,\nu}]$ denotes the commutator. 
The curvature is a measure for the nontrivial topology on $\mathcal{M}$, which 
manifests itself in the richness of holonomic transformations on the subspace 
$\mathcal{H}_{j}$. If the connection $A_{j}$ is irreducible, then the holonomy 
group coincides with the whole $\mathrm{U}(m_{j})$. In Ref.~\cite{HQC} it was 
proven that two generic loops in $\mathcal{M}$ are sufficient to generate a 
dense subset of the unitary group, that is, any element in $\mathrm{U}(m_{j})$ 
can be approximated to arbitrary precision by implementing a finite product 
sequence of these loops. Clearly, if $\mathcal{H}_{j}$ can be viewed as a 
multi-qubit code, then the irreducibility of the connection is equivalent to 
the notion of computational universality for $m_{j}\geq4$ \cite{Divi,Lloyd}. 

\section{Two-photon tripod holonomies}
\label{sec:example}

In this section, we illustrate the previous general scheme on the example of a 
photonic tripod represented in the two-photon Fock layer. For $M=3$ and $N=2$,
the relevant Fock states are
\begin{equation}
 \label{eq:F2}
 \begin{split}
 \mathcal{F}_{2}=\{&\ket{1100},\ket{0110},\ket{1010},\ket{0011},\ket{0101},\\
 &\ket{1001},\ket{2000},\ket{0200},\ket{0020},\ket{0002}\}.\\
 \end{split}
\end{equation}
In this Fock layer, the Hamiltonian from Eq.~(\ref{eq:M-pod}) gives rise to a 
four-dimensional dark subspace $\mathcal{H}_{\mathcal{D}}$, spanned by dark 
states $\{\ket{D_{a}(\bm{\kappa})}\}_{a=1}^{4}$ (details in App.~\ref{app:DS&BS}).
In the following, we show how to generate arbitrary $\mathrm{U}(4)$ transformations on 
$\mathcal{H}_{\mathcal{D}}$. To ensure this, we will enact a pair of 
noncommuting holonomies $W_{1}^{(\mathcal{D})}$ and $W_{2}^{(\mathcal{D})}$.
Let us choose local coordinates $\kappa_{i}=r_{i}\me^{\mi\theta_{i}}$, with 
$r_{i}\geq0$ and $\theta_{i}\in[0,2\pi)$ that parametrise the $\bm{\kappa}$-space 
$\mathcal{M}$. In the notation of Sec.~\ref{sec:HQC} this means $\lambda_{\mu}\in\{r_{i},\theta_{i}\}_{i=1}^{3}$.
For the generation of nontrivial holonomies, a certain richness of the control 
space $\mathcal{M}$ has to be provided. It turns out that, in our case, one 
needs at least three real coordinates (this holds even for all $N$ and $M$)
to ensure a nonvanishing curvature (\ref{eq:curvature}), that can give rise to 
a quantum holonomy.

For the first transformation we set $\kappa_{1}=0$, 
$\kappa_{2}=r_{2}\me^{\mi\theta_{2}}$, and $\kappa_{3}=r_{3}$.
Note, that there are indeed many existing schemes to realise (effective) 
complex coupling strengths, see e.g. Refs.~\cite{Esslinger,Cages}. Next, we derive the 
dark states with respect to the parametrisation $\{r_{2},\theta_{2},r_{3}\}$ of 
$\mathcal{M}$ as shown in Eq.~(\ref{eq:gauge}).
Due to the complex coupling one obtains connection 
coefficients with nonvanishing diagonal elements, that is
\begin{equation}
 A_{\theta_{2}}=\begin{pmatrix}
 0&0&0&0\\
 0&\frac{\mi r_{3}^2}{r_{2}^2+r_{3}^2}&0&0\\
 0&0&\frac{2\mi r_{3}^2}{r_{2}^2+r_{3}^2}&0\\
 0&0&0&\frac{\mi r_{2}^2}{r_{2}^2+r_{3}^2}\\
 \end{pmatrix}
\end{equation}
(all remaining components vanish),
written in the basis of dark states (view App.~\ref{app:DS&BS} for details) 
at an initial point $\bm{\kappa}_{0}$ in $\mathcal{M}$. 

A nonvanishing connection enables us to generate purely geometric rotations 
within the dark subspace. More precisely, traversing an arbitrary loop $\gamma$ 
in $\mathcal{M}$ results in the holonomy
\begin{equation}
\label{eq:U1}
W_{1}^{(\mathcal{D})}(\gamma)=\mathrm{exp}\left(\oint_{\gamma} A_{\theta_{2}}\mathrm{d}\theta_{2}\right)
\end{equation}
where path ordering becomes obsolete, due to the fact that there is only one relevant component.
The evaluation of the matrix exponential in Eq.~(\ref{eq:U1}) 
becomes quite simple due to the diagonal form of $A_{\theta_{2}}$.

For generating the second transformation, we activate all three couplings 
$\kappa_{1}=r_{1}$, $\kappa_{2}=r_{2}$, and $\kappa_{3}=r_{3}$.
A quite similar calculation as before reveals that the connection coefficients 
are given by $A_{r_{1}}=-\mi\sqrt{2}\zeta_{213}(\bm{r})\Sigma$ and $A_{r_{2}}=\mi\sqrt{2} \zeta_{123}(\bm{r})\Sigma$, 
while $A_{r_{3}}$ vanishes. Here we made use of the definitions
\begin{equation}
 \Sigma=\begin{pmatrix}
 0&1&0&0\\
 -1&0&1&0\\
 0&-1&0&0\\
 0&0&0&0\\
 \end{pmatrix},\quad \zeta_{ijk}(\bm{r})=\frac{r_{i}r_{k}}{(r_{i}^2+r_{j}^2)r},
\end{equation}
where $r=\sqrt{r_{1}^2+r_{2}^2+r_{3}^2}$.
Due to commuting connection coefficients, path ordering can be neglected again.
Hence, the corresponding matrix exponential can still be evaluated explicitly so that the holonomy reads
\begin{widetext}
\begin{equation}
\label{eq:U2}
W_{2}^{(\mathcal{D})}(\gamma)=\begin{pmatrix}
\cos^2(\phi_{0})&\sqrt{2}\sin(\phi_{0})\cos(\phi_{0})&\sin^2(\phi_{0})&0\\
-\sqrt{2}\sin(\phi_{0})\cos(\phi_{0})&\cos(2\phi_{0})&\sqrt{2}\sin(\phi_{0})\cos(\phi_{0})&0\\
\sin^2(\phi_{0})&-\sqrt{2}\sin(\phi_{0})\cos(\phi_{0})&\cos^2(\phi_{0})&0\\
0&0&0&1\\
\end{pmatrix},
\end{equation}
\end{widetext}
with 
\begin{equation}
 \phi_{0}(\gamma)=\oint_{\gamma}\zeta_{123}(\bm{r})\mathrm{d}r_2-\zeta_{213}(\bm{r})\mathrm{d}r_1
\end{equation}
being a geometric phase factor.
It turns out that the 
holonomies (\ref{eq:U1}) and (\ref{eq:U2}) do not commute 
for generic loops in $\mathcal{M}$.
Thus, it is shown how, in principle, any $\mathrm{U}(4)$ 
transformation on $\mathcal{H}_{\mathcal{D}}$ can be approximated arbitrarily 
well in terms of holonomies only. More generally, our photonic setup provides 
us, indeed, with the possibility to implement qudit states (i.e. 
$\mathcal{H}_{\mathcal{D}}\cong\mathbb{C}^{m}$) on which arbitrary 
$\mathrm{U}(m)$ holonomies act.

Further, note that in comparison to atomic physics, where a fivepod is 
necessary to obtain holonomic $\mathrm{U}(4)$ transformations, in photonic 
waveguide structures one only needs to design a tripod system. The optical 
setup has another advantage over the usual atomic scheme as we will illustrate 
in the following. In the representation of $\mathcal{F}_{2}$, the 
photonic structure gives also rise to a twofold degeneracy in the bright states 
$\ket{B_{1}^{(\pm)}(\bm{\kappa})},\ket{B_{2}^{(\pm)}(\bm{\kappa})}$,
which is absent in atomic $M$-pod systems.
Their respective eigenenergies are 
$\pm\sqrt{|\kappa_{1}|^2+|\kappa_{2}|^2+|\kappa_{3}|^2}$. It is straightforward 
to show that one can generate arbitrary $\mathrm{U}(2)$ manipulations on each 
of the eigenspaces $\mathcal{H}_{\mathcal{B}_{+}}$ and 
$\mathcal{H}_{\mathcal{B}_{-}}$. 
To be precise, let $\ket{B_{1}^{(\pm)}},\ket{B_{2}^{(\pm)}}$ be the bright 
states spanning $\mathcal{H}_{\mathcal{B}_{+}}$ and 
$\mathcal{H}_{\mathcal{B}_{-}}$, respectively. In the case of three real 
coupling parameters, the corresponding connections reads 
\begin{equation}
\label{eq:conn-coeff}
 A_{\pm,r_{1}}=-\mi \zeta_{312}(\bm{r}) \sigma_{y}^{(\pm)},\quad
 A_{\pm,r_{3}}=\mi \zeta_{132}(\bm{r}) \sigma_{y}^{(\pm)},
\end{equation}
while $A_{\pm,r_{2}}=0$.
Here, we defined
\begin{equation}
 \begin{split}
 \sigma_{y}^{(\pm)}&=\mi\ket{B_{2}^{(\pm)}(\bm{\kappa}_{0})} 
\bra{B_{1}^{(\pm)}(\bm{\kappa}_{0})}-\mi\ket{B_{1}^{(\pm)}(\bm{\kappa}_{0})}
\bra{B_{2}^{(\pm)}(\bm{\kappa}_{0})},
 \end{split}
\end{equation}
where we have chosen $\bm{\kappa}_{0}=(0,0,\kappa)$ 
($\kappa=\mathrm{const.}$) as the initial point at which the holonomy is 
generated. 
The general parameter-dependent form of the bright states is contained in App.~\ref{app:DS&BS}.
Fortunately, path ordering can be omitted, because all $A_{r_{i}}$ 
commute with one another, that means we found an Abelian substructure of the 
system. With the connection at hand, we are able to design the adiabatic 
holonomy 
\begin{equation}
\label{eq:gate}
W_{1}^{(\pm)}(\gamma)=\me^{\mi\phi_{1}(\gamma)\sigma_{y}^{(\pm)}}=
\begin{pmatrix}
\cos\phi_{1}&\sin\phi_{1}\\
-\sin\phi_{1}&\cos\phi_{1}\\
\end{pmatrix}
\end{equation}
with the geometric phase
\begin{equation}
\label{eq:phase}
\begin{split}
\phi_{1}(\gamma)&=\oint_{\gamma}\zeta_{132}(\bm{r})\mathrm{d}r_{3}-\zeta_{312}(\bm{r})\mathrm{d}r_{1},
\end{split}
\end{equation}
and the matrix in Eq.~(\ref{eq:gate}) being written in the basis
\begin{equation}
\label{eq:bright-basis}
 \begin{split}
 \ket{B_{1}^{(\pm)}(\bm{\kappa}_{0})}&=\frac{1}{\sqrt{2}}\left(\ket{1010}\pm\ket{1001}\right),\\
 \ket{B_{2}^{(\pm)}(\bm{\kappa}_{0})}&=\frac{1}{\sqrt{2}}\left(\ket{0110}\pm\ket{0101}\right).\\
 \end{split}
\end{equation}

A second holonomy can be obtained in a similar way. To ensure that the 
transformations do not commute, a complex coupling between one of the outer 
waveguides and the central one has to be implemented. Here, we set 
$\kappa_{1}=r_{1}\me^{\mi\theta_{1}}$. Moreover, let $\kappa_{2}=0$, so that we 
are only concerned with loops $\gamma$ described by coordinates 
$\{r_{1},\theta_{1},r_{3}\}$ on $\mathcal{M}$. At the point $\bm{\kappa}_{0}$, 
the holonomic unitary for this parametrisation takes the form 
$W_{2}^{(\pm)}=\me^{\mi\phi_{2}}\ket{B_{1}^{(\pm)}}\bra{B_{1}^{(\pm)}}
+\me^{\mi\tilde{\phi}_{2}}\ket{B_{2}^{(\pm)}}\bra{B_{2}^{(\pm)}}$, 
depending on the geometric phase factors
\begin{equation}
 \label{eq:gate-2}
\phi_{2}(\gamma)=\oint_{\gamma}\frac{r_{1}^2+2r_{3}^2}{2(r_{1}^2+r_{3}^2)}
\mathrm{d}\theta_{1},
\quad\tilde{\phi}_{2}(\gamma)=\oint_{\gamma}\frac{r_{1}^2}{2(r_{1}^2+r_{3}^2)}
\mathrm{d}\theta_{1},
\end{equation}
with $\gamma$ being another loop in $\mathcal{M}$. From here on it is easy to 
check that the transformations $W_{1}^{(\pm)}$ and $W_{2}^{(\pm)}$ do not 
commute, and thus the existence of a universal set of single-qubit gates on 
$\mathcal{H}_{\mathcal{B}_{\pm}}$ is verified.

Note that, unlike in the case of the dark states, the bright states accumulate 
not only a geometric phase but a dynamical phase as well. The latter one reads 
$\me^{\pm\mi\int_0^T\sqrt{|\kappa_1|^2+|\kappa_2|^2+|\kappa_3|^2}\mathrm{d}t}$. 
We should stress that such dynamical contributions do not possess the 
robustness and fault-tolerance of a purely holonomic quantum gate. However, it 
was proven in Ref.~\cite{Oreshkov3} that robust QC can be done efficiently on 
subsystems (different eigenspaces) of the total Hilbert space in a fully 
holonomic fashion. In Ref.~\cite{Oreshkov} it was further shown that such a 
computation can be made, in principle, completely fault-tolerant, by providing 
additional syndrome and gauge qubits.

\section{Remarks on the computational complexity of subsystems}
\label{sec:complexity}

As our theoretical proposal provides eigenspaces with arbitrarily large 
degrees of degeneracy, the question arises as to how efficient QC can be done 
on these spaces. We start by recalling that an $M$-pod filled with $N$ photons 
is described by $(N+M)!/(N!M!)$ orthogonal Fock states distributed over $2N+1$ 
separate eigenspaces. However, we have to note that generically an eigenspace 
$\mathcal{H}_{j}$, from the decomposition (\ref{eq:decomp}), does not need to 
support a proper multi-partite structure. By that we mean that there is no 
guarantee that one can decompose $\mathcal{H}_{j}$ into a product of 
single-qubit Hilbert spaces in any physically relevant way \cite{Zanardi}.
More formally speaking, the algebra of observables (here the $C^{*}$-algebra of 
bosonic creation and annihilation operators) does not inherit a tensor-product 
structure solely restricted to $\mathcal{H}_{j}$ \cite{Zanardi,ZanarLloyd}. 
This problem occurs frequently in the paradigm of HQC and has to be overcome to 
ensure a consistent labelling of logical qubits \cite{Pachos99}.

One possible solution to this difficulty might be to use the natural 
multi-partite structure induced by the Hamiltonian (\ref{eq:M-pod}). 
The total Hilbert space over which this observable acts can be decomposed with 
respect to the spatial modes of each waveguide, i.e. 
$\mathcal{H}=\mathrm{Span}(\{\otimes_{j}\ket{k}_{j}\}_{k=0}^{n_{j}})$ (recall 
that $\sum_{j}n_{j}=N$). From this point of view, we are able to implement a 
maximal number of $M+1$ qudits, with their dimension corresponding to the number 
of photons in the waveguide system. As simple this solution might seem at 
first sight, it leads to a rather subtle issue. In the scenario under 
investigation the generated holonomies may not act as a proper quantum gate 
solely within one of the eigenspaces, but rather on a logical quantum code 
$\mathcal{C}\subseteq\mathcal{H}$. A series of holonomies in different 
eigenspaces might then be needed to produce the desired transformation on the 
level of Fock states. 
Nevertheless, because one can generate any transformation 
on each of the respective eigenspaces, it may be well possible,
if the eigenspaces are large enough, to generate any linear optical computation
within a subspace of $\mathcal{H}$, which has a natural 
multi-partite structure inherited from the spatial-mode structure of the 
waveguides.

\subsection{Implementation of Two-Qubit States}
\label{ssec:2-qubit}

Let us clarify the above statements by a generic example. 
In the following, we will show how the two-photon tripod from 
Sec.~\ref{sec:example} serves as a sufficient setup for the implementation of 
two-qubit states. For that, recall that the first order bright subspaces 
$\mathcal{H}_{\mathcal{B}_{\pm}}$ at the point $\bm{\kappa}_{0}=(0,0,\kappa)$ 
are spanned by the states (\ref{eq:bright-basis}). Logical qubits are then 
defined with respect to the spatial mode structure of the waveguide network, 
viz. $\ket{0}_{\mathrm{L}}=\ket{1}_{1,3}\otimes\ket{0}_{2,4}$ and 
$\ket{1}_{\mathrm{L}}=\ket{0}_{1,3}\otimes\ket{1}_{2,4}$. 
With this definition at hand, the two-qubit states
\begin{equation}
 \label{eq:qubits}
 \begin{split}
 \ket{00}_{\mathrm{L}}&=\ket{1010},\qquad\ket{10}_{\mathrm{L}}=\ket{0110},\\
 \ket{01}_{\mathrm{L}}&=\ket{1001},\qquad\ket{11}_{\mathrm{L}}=\ket{0101},\\
 \end{split}
\end{equation}
lie completely within the quantum code 
$\mathcal{C}=\mathcal{H}_{\mathcal{B}_{+}}\oplus\mathcal{H}_{\mathcal{B}_{-}}$.
The labelling in Eq.~(\ref{eq:qubits}) preserves the underlying bipartite 
structure of the waveguides. Hence, we have a physical realisation of two-qubit 
states. After a (holonomic) quantum algorithm transformed an initial 
preparation $\ket{\psi_{\mathrm{in}}}$ into the desired answer 
$\ket{\psi_{\mathrm{out}}}$ of a computational problem, the output state can be 
measured by a set of photo detectors at the output facets of the waveguides.
Note that the bright states decompose into product states with respect to the 
bi-partition of the waveguides~(\ref{eq:qubits}),
\begin{equation}
 \label{eq:basis}
 \begin{split}
 \ket{B_{1}^{(\pm)}(\bm{\kappa}_{0})}&=\ket{0}_{\mathrm{L}}\otimes\ket{\pm},\\
 \ket{B_{2}^{(\pm)}(\bm{\kappa}_{0})}&=\ket{1}_{\mathrm{L}}\otimes\ket{\pm},\\
 \end{split}
\end{equation}
where $\ket{\pm}=(\ket{0}_{\mathrm{L}}\pm\ket{1}_{\mathrm{L}})/\sqrt{2}$ 
denotes the diagonal basis. With this explicit representation one can 
investigate how arbitrary holonomies act on the code $\mathcal{C}$ in terms of 
the qubits (\ref{eq:qubits}).

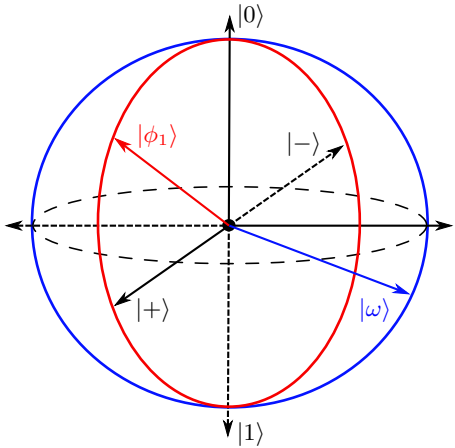
\begin{figure}[h]
\begin{tikzpicture}
\node at (0,0) {\includegraphics[width=6cm]{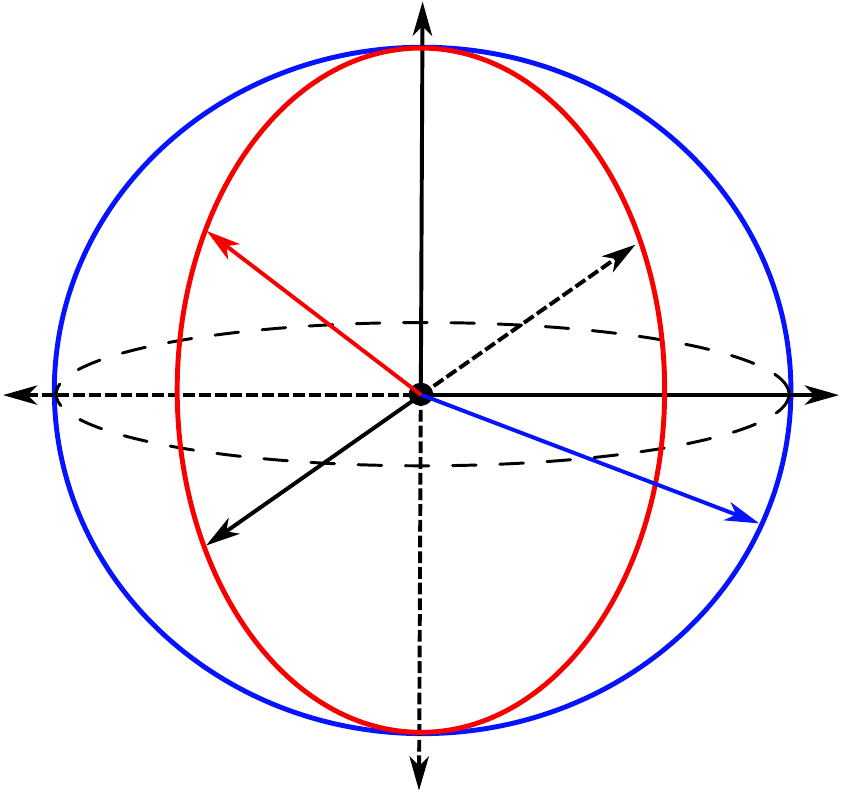}}; \node at (-0.94,1.21) {\textcolor{red}{$\ket{\phi_{1}}$}}; \node at (1.95,-1.1) {\textcolor{blue}{$\ket{\omega}$}};
\node at (0.3,2.8) {$\ket{0}$}; \node at (0.3,-2.75) {$\ket{1}$};
\node at (-1,-1.1) {$\ket{+}$}; \node at (1.,1.1) {$\ket{-}$};
\end{tikzpicture}
\caption{\label{fig:Bloch} Bloch sphere representation of the states $\ket{\phi_{1}}$ and $\ket{\omega}$. 
Manipulating the geometric phase factor $\phi_{1}$ results in a motion along the red great circle. In comparison, a change in the dynamical part $\omega$ moves the state along the blue great circle.
By combining both transformations any state on the the Bloch sphere can be reached.}
\end{figure}

For the purpose of illustration, let us focus on a benchmark holonomy. In 
Sec.~\ref{sec:example} we explained that the connections over 
$\mathcal{H}_{\mathcal{B}_{+}}$ and $\mathcal{H}_{\mathcal{B}_{-}}$ are 
irreducible. It thus follows that, by adiabatically varying the 
Hamiltonian (\ref{eq:M-pod}) along a suitable loop in $\mathcal{M}$, we are 
able to apply the gate [cf. Eq.~(\ref{eq:full-holo})]
\begin{equation}
 \label{eq:qubit-gate}
 U(\omega_{\pm},\gamma)=\me^{\mi\omega_{+}}W^{(+)}(\gamma)\oplus \me^{\mi\omega_{-}}W^{(-)}(\gamma)
\end{equation} 
to the qubits (\ref{eq:qubits}), where $W^{(+)}$ and $W^{(-)}$ are now 
arbitrary holonomies acting within each subspace on the states (\ref{eq:basis}).
Recall that $\omega_{\pm}$ denote the integrals over the eigenenergies of 
$\mathcal{H}_{\mathcal{B}_{+}},\mathcal{H}_{\mathcal{B}_{-}}$, respectively.
In particular, it holds that $\omega_{\pm}=\pm\omega$ with $\omega>0$.
For concreteness, let us consider the unitaries from Eq.~(\ref{eq:gate}).
Then, a composite holonomy $U_{1}(\omega,\phi)$ [cf. Eq.~(\ref{eq:qubit-gate})] 
acts on the computational basis~(\ref{eq:qubits}) according to the truth table
\begin{equation}
 \label{eq:truth-table}
 \begin{split}
 \ket{00}_{\mathrm{L}}&\mapsto\ket{\phi_{1}}\otimes\ket{\omega},\qquad
 \ket{10}_{\mathrm{L}}\mapsto\ket{\overline{\phi}_{1}}\otimes\ket{\omega},\\ 
\ket{01}_{\mathrm{L}}&\mapsto\ket{\phi_{1}}\otimes\ket{\overline{\omega}},\qquad
\ket{11}_{\mathrm{L}}\mapsto\ket{\overline{\phi}_{1}}\otimes\ket{\overline{
\omega}},\\
 \end{split}
\end{equation}
where we introduced the $\phi_{1}$-dependent states
\begin{equation}
 \label{eq:phi-states}
 \begin{split}
 \ket{\phi_{1}}&=\cos\phi_{1}\ket{0}_{\mathrm{L}}
 -\sin\phi_{1}\ket{1}_{\mathrm{L}},\\
 \ket{\overline{\phi}_{1}}&=\cos\phi_{1}\ket{1}_{\mathrm{L}}
 +\sin\phi_{1}\ket{0}_{\mathrm{L}},\\
 \end{split}
\end{equation}
together with the $\omega$-dependent states
\begin{equation}
 \label{eq:omega-states}
 \begin{split}
 \ket{\omega}&=\cos\omega\ket{0}_{\mathrm{L}}
 +\mi\sin\omega\ket{1}_{\mathrm{L}},\\
 \ket{\overline{\omega}}&=\cos\omega\ket{1}_{\mathrm{L}}
 +\mi\sin\omega\ket{0}_{\mathrm{L}}.\\
 \end{split}
\end{equation}
The latter describe a dynamical superposition of the vacuum and the one-photon 
Fock state localized in the third or fourth waveguide,
while the former are of purely geometric origin and distribute over the 
first two modes. By designing a plaquette $\square$ in $\mathcal{M}$ such that 
$\phi_{1}(\square)=-\pi/4$ (details in App.~\ref{app:loops}) the logical qubits 
obey the transformation $XH\otimes\Omega(\omega)$, where a Hadamard gate 
$H=\ket{0}_{\mathrm{L}}\bra{+}+\ket{1}_{\mathrm{L}}\bra{-}$ acts on the first 
qubit (first and second waveguide) followed by the bit-flip gate 
$X=\ket{0}_{\mathrm{L}}\bra{1}_{\mathrm{L}}
+\ket{1}_{\mathrm{L}}\bra{0}_{\mathrm{L}}$, while simultaneously the gate 
$\Omega(\omega)=\ket{\omega}\bra{0}_{\mathrm{L}}+\ket{\overline{\omega}}\bra{1}_{\mathrm{L}}$ 
acts on the second qubit (third and fourth waveguide). 
The gate $\Omega$ parametrises a great circle traversing through the poles 
($\ket{0}_{\mathrm{L}}$ and $\ket{1}_{\mathrm{L}}$) of the Bloch sphere (cf. 
Fig.~\ref{fig:Bloch}). In comparison, the action on the first qubit has no 
inherent dynamical contribution and is therefore robust towards experimental 
imperfections that undermine the plaquette $\square$. 

Designing another loop $\gamma$ such that $\phi_{1}(\gamma)=\pi/2$ (cf. 
App.~\ref{app:loops}) creates the quantum circuit $\mi Y\otimes\Omega(\omega)$, 
with $Y=\mi(\ket{1}_{\mathrm{L}}\bra{0}_{\mathrm{L}}
-\ket{0}_{\mathrm{L}}\bra{1}_{\mathrm{L}})$ being the Pauli-$Y$ gate. 
If the experimenter is able to change the rate with which the path is 
traversed, this will not influence the quantum holonomic gate (as long as 
adiabaticity holds) but can realise a desired transformation on the second qubit 
in terms of a dynamical gate.

A different quantum gate $U_{2}(\omega,\gamma)$ is obtained by inserting 
the holonomy $W_{2}^{(\pm)}(\gamma)$ (recall Sec.~\ref{sec:example}) into 
Eq.~(\ref{eq:qubit-gate}). This gate obeys the truth table
\begin{equation}
 \label{eq:truth-table-3}
 \begin{split}
 \ket{00}_{\mathrm{L}}&\mapsto\me^{\mi\phi_{2}}\ket{0}_{\mathrm{L}}\otimes
\ket{\omega},\quad\ket{10}_{\mathrm{L}}\mapsto\me^{\mi\tilde{\phi}_{2}}
\ket{1}_{\mathrm{L}}\otimes\ket{\omega},\\
 \ket{01}_{\mathrm{L}}&\mapsto\me^{\mi\phi_{2}}\ket{0}_{\mathrm{L}}\otimes
\ket{\overline{\omega}},\quad
\ket{11}_{\mathrm{L}}\mapsto\me^{\mi\tilde{\phi}_{2}}
\ket{1}_{\mathrm{L}}\otimes\ket{\overline{\omega}},\\
 \end{split}
\end{equation}
with the path-dependent geometric phases $\phi_{2}$ and $\tilde{\phi}_{2}$ from 
Eq.~(\ref{eq:gate-2}). The unitary (\ref{eq:truth-table-3}) corresponds to a 
product of single qubit gates, that is, $U_{2}=P\otimes\Omega$, with 
$P=\me^{\mi\phi_{2}}\ket{0}_{\mathrm{L}}\bra{0}_{\mathrm{L}}
+\me^{\mi\tilde{\phi}_{2}}\ket{1}_{\mathrm{L}}\bra{1}_{\mathrm{L}}$ being a 
phase gate. The quantum gates $U_{1}$ and $U_{2}$ in general do not commute.
From Fig.~\ref{fig:Bloch} we can conclude that the gates we provided form a 
dense subset of the unitary group 
$\mathrm{U}(2)\otimes\mathrm{U}(2)\subset\mathrm{U}(4)$ and thus, we are,
in principle, able to approximate any of its elements to arbitrary precision.

\subsection{Implementation of Three-Qubit States}
\label{ssec:3-qubit}

For the preparation of larger code words than two-qubit states we need to 
enlarge the number of states on which to perform holonomic computation, that 
is, we need to prepare larger degenerate eigenspaces. Because encoded 
information lies in a proper subspace $\mathcal{C}$ of the total Hilbert space 
$\mathcal{H}$, the quantum code is most generally a subspace reaching over 
several eigenspaces of $\mathcal{H}$. Let us illustrate this point by 
constructing three-qubit states, i.e. we have 
$\mathcal{C}\cong(\mathbb{C}^2)^{\otimes^3}\cong\mathbb{C}^8$.
Therefore, one has to inject another photon into the optical tripod from 
Sec.~\ref{sec:example}, that is, we study the tripod in the three-photon Fock 
layer. Analogously to Sec.~\ref{sec:example}, one can show that, in principle, 
any unitary transformation can be carried out over the first-order bright 
subspaces $\mathcal{H}_{\mathcal{B}_{+}}$ and $\mathcal{H}_{\mathcal{B}_{-}}$ in 
terms of quantum holonomies. Both eigenspaces are fourfold degenerate and are 
spanned as $\mathcal{H}_{\mathcal{B}_{\pm}}=
\mathrm{Span}(\{\ket{B_{a}^{(\pm)}}\}_{a=1}^4)$. 
In addition, we take the nondegenerate third-order bright subspaces 
$\mathcal{H}_{\mathcal{B}_{+}}^{(3)}$ and $\mathcal{H}_{\mathcal{B}_{-}}^{(3)}$ 
into account. Subsequently, the eight-dimensional quantum code $\mathcal{C}$ 
will be a proper subspace of the ten-dimensional Hilbert space 
$\mathcal{H}_{\mathcal{B}_{+}}\oplus\mathcal{H}_{\mathcal{B}_{-}}\oplus
\mathcal{H}_{\mathcal{B}_{+}}^{(3)}\oplus\mathcal{H}_{\mathcal{B}_{-}}^{(3)}$
that preserves the tri-partite (spatial-mode) structure of the three outer 
waveguides. 

We found a consistent labelling to be 
\begin{equation}
 \label{eq:labelling}
 \begin{split}
 \ket{000}_{\mathrm{L}}=\ket{0003},&\quad\ket{100}_{\mathrm{L}}=\ket{3000},\\
 \ket{010}_{\mathrm{L}}=\ket{0201},&\quad\ket{001}_{\mathrm{L}}=\ket{0021},\\
 \ket{110}_{\mathrm{L}}=\ket{1200},&\quad\ket{101}_{\mathrm{L}}=\ket{1020},\\
 \ket{011}_{\mathrm{L}}=\ket{0111},&\quad\ket{111}_{\mathrm{L}}=\ket{1110}.\\
 \end{split}
\end{equation}
In Eq.~(\ref{eq:labelling}), the logical zero corresponds to the vacuum state,
and the logical one is encoded when at least one photon impinges onto the 
detector in the respective waveguide. The qubits (Fock states) in 
Eq.~(\ref{eq:labelling}) can be expressed solely through bright states in 
$\mathcal{H}_{\mathcal{B}_{+}}\oplus\mathcal{H}_{\mathcal{B}_{-}}\oplus
\mathcal{H}_{\mathcal{B}_{+}}^{(3)}\oplus\mathcal{H}_{\mathcal{B}_{-}}^{(3)}$, 
hence at the point $\bm{\kappa}_{0}=(\kappa,0,0)$ we have 
\begin{widetext}
\begin{equation}
 \begin{split}
\ket{000}_{\mathrm{L}}&=\frac{1}{2\sqrt{2}}\left(\ket{B^{(+3)}(\bm{\kappa}_{0})}
+\ket{B^{(-3)}(\bm{\kappa}_{0})}\right)+\frac{\sqrt{3}}{2\sqrt{2}}
\left(\ket{B_{1}^{(+)}(\bm{\kappa}_{0})}+\ket{B_{1}^{(-)}(\bm{\kappa}_{0})}
\right),\\
\ket{100}_{\mathrm{L}}&=\frac{1}{2\sqrt{2}}\left(\ket{B^{(+3)}(\bm{\kappa}_{0})}
-\ket{B^{(-3)}(\bm{\kappa}_{0})}\right)-\frac{\sqrt{3}}{2\sqrt{2}}
\left(\ket{B_{1}^{(+)}(\bm{\kappa}_{0})}-\ket{B_{1}^{(-)}(\bm{\kappa}_{0})}
\right),\\
 \end{split}
\end{equation}
\begin{equation}
 \begin{split}
\ket{010}_{\mathrm{L}}&=\frac{1}{\sqrt{2}}
\left(\ket{B_{4}^{(+)}(\bm{\kappa}_{0})}
+\ket{B_{4}^{(-)}(\bm{\kappa}_{0})}\right),\qquad\ket{001}_{\mathrm{L}}=
\frac{1}{\sqrt{2}}\left(\ket{B_{2}^{(+)}(\bm{\kappa}_{0})}
+\ket{B_{2}^{(-)}(\bm{\kappa}_{0})}\right),\\
 \ket{110}_{\mathrm{L}}&=\frac{1}{\sqrt{2}}
 \left(\ket{B_{4}^{(+)}(\bm{\kappa}_{0})}-\ket{B_{4}^{(-)}(\bm{\kappa}_{0})}
\right),\qquad
\ket{101}_{\mathrm{L}}=\frac{1}{\sqrt{2}}\left(\ket{B_{2}^{(+)}
(\bm{\kappa}_{0})}-\ket{B_{2}^{(-)}(\bm{\kappa}_{0})}\right),\\
 \ket{011}_{\mathrm{L}}&=\frac{1}{\sqrt{2}}
 \left(\ket{B_{3}^{(+)}(\bm{\kappa}_{0})}+\ket{B_{3}^{(-)}(\bm{\kappa}_{0})}
\right),\qquad\ket{111}_{\mathrm{L}}=\frac{1}{\sqrt{2}}\left(\ket{B_{3}^{(+)}
(\bm{\kappa}_{0})}-\ket{B_{3}^{(-)}(\bm{\kappa}_{0})}\right).\\
 \end{split}
\end{equation}
\end{widetext}

Note that there are two remaining states $\ket{1002}$ and $\ket{2001}$ in 
$\mathcal{H}_{\mathcal{B}_{+}}\oplus\mathcal{H}_{\mathcal{B}_{-}}\oplus
\mathcal{H}_{\mathcal{B}_{+}}^{(3)}\oplus\mathcal{H}_{\mathcal{B}_{-}}^{(3)}$ 
that can, in principle, introduce an over-labelling of logical states. 
When the experimenter is just observing if a detector, placed at the output 
facet of the first waveguide, did or did not click, the states can be 
mistaken as the qubit $\ket{100}_{\mathrm{L}}$. To avoid this undesirable 
phenomenom, we do not choose arbitrary holonomies on 
$\mathcal{H}_{\mathcal{B}_{+}}\oplus\mathcal{H}_{\mathcal{B}_{-}}\oplus\mathcal{
H}_{\mathcal{B}_{+}}^{(3)}\oplus\mathcal{H}_{\mathcal{B}_{-}}^{(3)}$, 
but those which map the code $\mathcal{C}$, i.e. the qubits from 
Eq.~(\ref{eq:labelling}), onto itself.

\section{Conclusion}
\label{sec:conclusions}

In this article, we have presented a theoretical proposal for the 
implementation of arbitrary $\mathrm{U}(m)$ transformations, to satisfactory 
precisions, by holonomic means. Our photonic setup consisted solely of 
directional couplers which were arranged as an $M$-pod system. We found that,
upon injecting additional photons into the waveguide system, the degeneracy of 
eigenspaces scaled drastically. 
In comparison to atomic $M$-pod systems, 
where a linear increase in degeneracy can only be observed in the dark subspace, 
our setup can produce a nonlinear increase in the degeneracy of each eigenspace.

For the special scenario of the two-photon optical tripod, the associated 
connections revealed that arbitrary geometric transformations can be designed on 
the eigenspaces of the system. This was explicitly shown for the group 
$\mathrm{U}(4)$.
We have shown how one can use the spatial mode structure of the 
waveguides to define a consistent labelling of logical qubits. 
This was explicitly demonstrated for the case of two-qubit states. 
Our scheme provides the utility of implementing robust quantum gates in terms 
of a composite holonomy generated over the twofold degenerate bright subspaces 
of the system. Moreover, we showed how three-qubit states can be labelled on the 
optical tripod by injecting a third photon, thus illustrating that our photonic 
scheme is scalable in terms of providing additional qubits.

Our article paves the way for an experimental study of large holonomy groups 
and the transformation behaviour of bosonic Fock states under their action. 
It shows that the problem of having no natural multi-partite structure on the 
computational eigenspaces can be overcome by considering subspaces so large, 
that a natural multi-partite quantum code can be realised within them. 

\acknowledgments
Financial support by the Deutsche Forschungsgemeinschaft (DFG SCHE 612/6-1) 
is gratefully acknowledged.

\appendix

\begin{widetext}
\section{Dark States and Bright States for the Two-Photon Tripod}
\label{app:DS&BS}

In the following, explicit formulas for the dark states and first order bright states of the two-photon tripod are given. 
From these states, one can directly calculate the respective connections according to Eq.~(\ref{eq:gauge}). 
Recall the first parametrisation used for the dark subspace holonomy, with $\kappa_{1}=0$, $\kappa_{2}=r_{2}\me^{\mi\theta_{2}}$,
and $\kappa_{3}=r_{3}$ with $r_{i}\geq0$ and $\theta_{i}\in[0,2\pi)$.
Under this parametrisation the (orthogonalised) dark states take the form 
\begin{equation}
 \begin{split}
 \ket{D_{1}}&=\ket{2000},\\
 \ket{D_{2}}&=\frac{1}{\rho_{23}}\left(r_{2}\ket{1010}-r_{3}\me^{\mi\theta_{2}}\ket{1100}\right),\\
 \ket{D_{3}}&=\frac{1}{\rho_{23}^2}\left(r_{3}^2\me^{2\mi\theta_{2}}\ket{0200}-\sqrt{2}r_{2}r_{3}\me^{\mi\theta_{2}}\ket{0110}+r_{2}^2\ket{0020}\right),\\
 \ket{D_{4}}&=\frac{1}{\sqrt{2}\rho_{23}^2}\left(r_{2}^2\me^{2\mi\theta_{2}}\ket{0200}+r_{3}^2\ket{0020}\right)+\frac{r_{2}r_{3}}{\rho_{23}^2}\me^{\mi\theta_{2}}\ket{0110}-\frac{1}{\sqrt{2}}\ket{0002},\\
 \end{split}
\end{equation}
with $\rho_{ij}=\sqrt{r_{i}^2+r_{j}^2}$. 
Next, for the real-valued coordinates $\kappa_{1}=r_{1}$, $\kappa_{2}=r_{2}$, and $\kappa_{3}=r_{3}$, the dark states read
\begin{equation}
 \begin{split}
 \ket{D_{1}}&=\frac{1}{\rho_{12}^2}\left(r_{2}^2\ket{2000}-\sqrt{2}r_{1}r_{2}\ket{1100}+r_{1}^2\ket{0200}\right),\\
 \ket{D_{2}}&=\frac{\sqrt{2}r_{1}r_{2}r_{3}}{\rho_{12}^2 r}\left(\ket{0200}-\ket{2000}\right)+\frac{1}{r}\left(r_{2}\ket{1010}-r_{1}\ket{0110}\right)+\frac{(r_{1}^2-r_{2}^2)r_{3}}{\rho_{12}^2 r}\ket{1100},\\
 \ket{D_{3}}&=\frac{r_{3}^2}{\rho_{12}^2 r^2}\left(r_{1}^2\ket{2000}+r_{2}^2\ket{0200}\right)+\frac{\sqrt{2}r_{3}}{r^2}\left(\frac{r_{1}r_{2}r_{3}}{\rho_{12}^2}\ket{1100}-r_{1}\ket{1010}-r_{2}\ket{0110}\right)
  +\frac{\rho_{12}^2}{r^2}\ket{0020},\\
 \ket{D_{4}}&=\frac{1}{\sqrt{2}r^2}\left(r_{1}^2\ket{2000}+r_{2}^2\ket{0200}+r_{3}^2\ket{0020}\right)+\frac{1}{r^2}\left(r_{1}r_{2}\ket{1100}+r_{1}r_{3}\ket{1010}+r_{2}r_{3}\ket{0110}\right)-\frac{1}{\sqrt{2}}\ket{0002},\\
 \end{split}
\end{equation}
where we defined $r=\sqrt{r_{1}^2+r_{2}^2+r_{3}^2}$.
From these two sets of states the adiabatic connection and subsequently the holonomies~(\ref{eq:U1}) and~(\ref{eq:U2}) were computed. 

We now turn to the first-order bright states, which were used in Subsection~\ref{ssec:2-qubit} to define logical two-qubit states. 
To obtain the geometric phase factors in Eq.~(\ref{eq:gate-2}) we chose local coordinates $\kappa_{1}=r_{1}\me^{\mi\theta_{1}}$, $\kappa_{2}=0$, and $\kappa_{3}=r_{3}$. 
The first order bright states in $\mathcal{H}_{\mathcal{B}_{+}}$ and  $\mathcal{H}_{\mathcal{B}_{-}}$ are found to be 
\begin{equation}
 \begin{split}
 \ket{B_{1}^{(\pm)}}&=\frac{r_{1}r_{3}}{\rho_{13}^2}\left(\me^{2\mi\theta_{1}}\ket{2000}-\ket{0020}\right)\pm\frac{1}{\sqrt{2}\rho_{13}}\left(r_{3}\me^{\mi\theta_{1}}\ket{1001}-r_{1}\ket{0011}\right)
                      +\frac{r_{3}^2-r_{1}^2}{\sqrt{2}\rho_{13}^2}\me^{\mi\theta_{1}}\ket{1010},\\
 \ket{B_{2}^{(\pm)}}&=\frac{1}{\sqrt{2}\rho_{13}}\left(r_{3}\ket{0110}+r_{1}\me^{\mi\theta_{1}}\ket{1100}\right)\pm\frac{1}{\sqrt{2}}\ket{0101}.\\
 \end{split}
\end{equation}
For the second transformation the couplings were $\kappa_{1}=r_{1}$, $\kappa_{2}=r_{2}$, and $\kappa_{3}=r_{3}$. 
The bright states for this case are 
\begin{equation}
 \begin{split}
 \ket{B_{1}^{(\pm)}}&=\frac{r_{1}r_{3}}{r\rho_{13}}\left(\ket{2000}-\ket{0020}\right)+\frac{r_{2}}{\sqrt{2}r\rho_{13}}\left(r_{3}\ket{1100}-r_{1}\ket{0110}\right)
 \pm\frac{1}{\sqrt{2}\rho_{13}}\left(r_{3}\ket{1001}-r_{1}\ket{0011}\right)+\frac{r_{3}^2-r_{1}^2}{\sqrt{2}r\rho_{13}}\ket{1010},\\
 \ket{B_{2}^{(\pm)}}&=\frac{\rho_{13}}{r}\left(\frac{r_{2}}{r}\ket{0200}\pm\frac{1}{\sqrt{2}}\ket{0101}\right)-\frac{r_{2}}{\rho_{13}r^2}\left(r_{1}^2\ket{2000}+\sqrt{2}r_{1}r_{3}\ket{1010}+r_{3}^2\ket{0020}\right)\\
 &\hspace{0.5cm}\mp\frac{r_{2}}{\sqrt{2}r\rho_{13}}\left(r_{1}\ket{1001}+r_{3}\ket{0011}\right)+\frac{r_{1}^2-r_{2}^2+r_{3}^2}{\sqrt{2}\rho_{13}r^2}\left(r_{1}\ket{1100}+r_{3}\ket{0110}\right).\\
 \end{split}
\end{equation}
From the above states we were able to compute the holonomy~(\ref{eq:gate}). 
Moreover, one can easily check that in the limit $\bm{\kappa}\rightarrow\bm{\kappa}_{0}=(0,0,\kappa)$ the basis~(\ref{eq:bright-basis}) is reproduced.

\end{widetext}

\section{Dimension of Subspaces}
\label{app:subSpaceDim}

The dimensions of the subspaces in an $M$-pod with $N$ photons injected can be determined as follows.
In case of only one photon or excitation the same result applies as in atomic physics~\cite{Dalibard}, i.e. in an $M$-pod term scheme there is one negative bright state, one positive and $M-1$ dark states.
This means that the Hamiltonian in Eq.~\eqref{eq:M-pod} can be rewritten with new bosonic mode operators $b_-, d_1, \dots, d_{M-1}, b_+$ as
\begin{align}
 H = - \varepsilon  b^\dagger_- b_- + \varepsilon b^\dagger_+ b_+,
\end{align}
where only the bright modes occur because of the zero eigenvalue of the dark modes.
The advantage of this representation of $H$ is that we can now simply turn to a Fock state notation when considering the case of $N$ photons.
The eigenstates of the Hamiltonian can then be given by the number of photons $n_-$ in the negative bright mode, the number of photons $n_+$ in the positive bright mode,
and $n_\mathcal{D}$ as the number of photons in the $M-1$ dark modes.
The eigenvalue equation of these Fock states is
\begin{align}
 H | n_-, n_\mathcal{D}, n_+ \rangle  = \varepsilon (n_+ - n_-) | n_-, n_\mathcal{D}, n_+ \rangle 
\end{align}
with $n_- + n_\mathcal{D} + n_+ = N$.

Counting the number of dark states in the $N$ photon case then amounts to counting the number of possibilities to distribute $N$ photons so that the eigenvalue $\varepsilon (n_+ - n_-)$ becomes zero.
Clearly, this is the case when either all $N$ photons are distributed over the $M-1$ dark modes, or when an equal number of photons are in the positive and negative bright modes with the rest in the dark modes.
For example, there are ${ {N + M - 2} \choose {2 n}}$ ways of distributing all photons over the dark modes.
Next, one can have one photon in each positive and negative bright mode, thus ${ {N-2 + M - 2} \choose {N-2}}$ ways remain to distribute the rest of the photons over the dark modes.
This continues until all photons are equally distributed over the positive and negative bright modes.
However, there are two distinct cases, $N$ odd or even, for which one finds two formulas for the total number of dark states, i.e.
\begin{align}
 d(N,M) = \begin{cases}
           \sum_{n=1}^{N/2} { {2 n + M - 2} \choose {2 n}} , & \text{if $N$ even,} \\
           \sum_{n=0}^{(N-1)/2} { {2 n + 1 + M - 2} \choose {2 n + 1}}, & \text{if $N$ odd.}
          \end{cases}
\end{align}

When counting the number of bright states with energy $\pm k \varepsilon$, one first has to put $k$ photons in either the negative or positive bright mode, and then distribute the rest as if to create a dark state.
Thus, for bright states with energy $\pm k \varepsilon$ there are $d(N-k,M)$ possibilities.

\section{Computation of Non-Abelian Geometric Phases}
\label{app:loops}

Here, we shall give an explicit way to obtain the geometric phases $\phi_{1}(\gamma)$ that implement the desired quantum gates in Subsection~\ref{ssec:2-qubit}. 
We recall from Sec.~\ref{sec:example} that the holonomies in Eq.~(\ref{eq:gate}), which act
on the bright subspaces $\mathcal{H}_{\mathcal{B}_{+}}$ and $\mathcal{H}_{\mathcal{B}_{-}}$ respectively, 
are completely determined by the scalar $\phi_{1}(\gamma)$ from Eq.~(\ref{eq:phase}). 
The line integral on which the phase factor depends can be replaced by a surface integral using Stokes' Theorem~\cite{Nakahara}
\begin{equation}
 \oint_{\gamma}A_{\pm}=\int_{\mathcal{S}}F_{\pm}=\mi\phi_{1}(\mathcal{S})\sigma_{y}^{(\pm)},
\end{equation}
where $\mathcal{S}$ is the area in the $D$-dimensional control space $\mathcal{M}$, which is enclosed by the loop $\gamma$. 
Given the connection coefficients from Eq.~(\ref{eq:conn-coeff}), one readily obtains the curvature~(\ref{eq:curvature}) with respect to $\mathcal{H}_{\mathcal{B}_{\pm}}$, namely
\begin{equation}
 \label{eq:curv-coeff}
 \begin{split}
 F_{\pm,r_{1}r_{2}}&=\mi r_{3}/r^3\sigma_{y}^{(\pm)},\\
 F_{\pm,r_{1}r_{3}}&=-\mi r_{2}/r^3\sigma_{y}^{(\pm)},\\
 F_{\pm,r_{2}r_{3}}&=\mi r_{1}/r^3\sigma_{y}^{(\pm)}.\\
 \end{split}
\end{equation}
Note that in Eq.~(\ref{eq:curv-coeff}) all components of the curvature commute so that not only is path ordering obsolete, but the commutator in Eq.~(\ref{eq:curvature}) vanishes.

As the precise form of the generating loop does not matter, one might as well design a simple plaquette 
\begin{equation}
\label{eq:square}
 \square(\alpha,\beta)=\left\{\bm{r}\in\mathbb{R}^3\,\bigr|\,r_{1}\in[0,\alpha],\,r_{2}\in[0,\beta],\,r_{3}=\kappa\right\},
\end{equation}
that is, restricting oneself to the $(r_{1},r_{2})$-plane at $r_{3}=\kappa=\mathrm{const}$. 
In this context, $\alpha$ and $\beta$ determine the area enclosed by $\square$, such that the desired geometric phase factor can be attained.
Under the constraints given by the plaquette~(\ref{eq:square}), the integration over an oriented surface reduces to $\iint_{\square}F_{\pm,r_{1}r_{2}}\mathrm{d}r_{1}\mathrm{d}r_{2}$.
Hence, the relevant phase factor becomes
\begin{equation}
 \phi_{1}(\square)=\int_{0}^{\beta}\int_{0}^{\alpha}\frac{\kappa\mathrm{d}r_{1}\mathrm{d}r_{2}}{\sqrt{r_{1}^2+r_{2}^2+\kappa^2}^3}.
\end{equation}
Fortunately, the integration can be performed analytically so that we obtain
\begin{equation}
 \label{eq:result}
 \phi_{1}(\alpha,\beta)=\mathrm{arctan}\left(\frac{\alpha\beta}{\kappa\sqrt{\alpha^2+\beta^2+\kappa^2}}\right).
\end{equation}
By appropriately choosing $\alpha>\kappa$ and $\beta=\kappa\sqrt{(\alpha^2+\kappa^2)/(\alpha^2-\kappa^2)}$, the phase factor~(\ref{eq:result}) can be set to $\phi_{1}(\alpha,\beta)=-\pi/4$,
which implements the gate $XH$ on the first qubit (cf. Subsection.~\ref{ssec:2-qubit}).

Next, we show how to realise the holonomic gate $\mi Y$ also discussed in Subsection~\ref{ssec:2-qubit}.
In this case, it turns out to be suitable to parametrise the $(\kappa_{1},\kappa_{2})$-plane as $\kappa_{1}=\kappa\sin\varphi\cos\vartheta$ and $\kappa_{2}=\kappa\sin\varphi\sin\vartheta$,
with $\varphi,\vartheta\in[0,2\pi)$, which can involve negative couplings for certain values of $\varphi$ and $\vartheta$. 
The surface integral transforms accordingly to
\begin{equation}
 \label{eq:sur-int}
 \phi_{1}(\mathcal{S})=\iint_{\mathcal{S}}\frac{\kappa}{\sqrt{\kappa_{1}^2+\kappa_{2}^2+\kappa^2}^3}\frac{\partial(\kappa_{1},\kappa_{2})}{\partial(\varphi,\vartheta)}\mathrm{d}\varphi\mathrm{d}\vartheta,
\end{equation}
where the Jacobian is 
\begin{equation}
 \frac{\partial(\kappa_{1},\kappa_{2})}{\partial(\varphi,\vartheta)}=\kappa^2\sin\varphi\cos\varphi>0
\end{equation}
for $\varphi\in(0,\pi/2)$.
Direct integration of~(\ref{eq:sur-int}) gives
\begin{equation}
\begin{split}
 \phi_{1}(\mathcal{S})&=\int_{\vartheta_{0}}^{\vartheta_{1}}\int_{\varphi_{0}}^{\varphi_{1}}\frac{\sin\varphi\cos\varphi}{\sqrt{1+\sin^2\varphi}}\mathrm{d}\varphi\mathrm{d}\vartheta,\\
 &=\left(\sqrt{\sin^2\varphi_{1}+1}-\sqrt{\sin^2\varphi_{0}+1}\right)\left[\vartheta_{1}-\vartheta_{0}\right].\\
\end{split} 
\end{equation}
In order to implement the desired gate, we shall design the geometric phase~(\ref{eq:sur-int}) as $\phi_{1}(\mathcal{S})=\pi/2$ for which
we can choose $\varphi_{0}=0$, $\varphi_{1}=\pi/2$, and $\vartheta_{1}-\vartheta_{0}=\pi/(2\sqrt{2}-2)$.

\end{document}